\documentclass[prc,aps,showpacs]{revtex4}
\usepackage{graphicx}
\begin{document}

\title{Light Baryon Resonances: Restrictions and Perspectives}

\author{Ya.~I.~Azimov}
\email{azimov@pa1400.spb.edu}
\affiliation{Petersburg Nuclear Physics Institute,
Gatchina \\
St.~Petersburg 188300, Russia}

\author{R.~A.~Arndt}
\email{arndt@reo.ntelos.net}
\author{I.~I.~Strakovsky}
\email{igor@gwu.edu}
\author{R.~L.~Workman}
\email{rworkman@gwu.edu}
\affiliation{Center for Nuclear Studies, Department of
        Physics, \\
        The George Washington University, Washington,
        D.C. 20052, U.S.A.}

%%%%%%%%%%%%%%%%%%%%%%%%%%%%%%%%%%%%%%%%%%%%%%%%%%%%%%%%%% 
%%%   Abstract
%%%%%%%%%%%%%%%%%%%%%%%%%%%%%%%%%%%%%%%%%%%%%%%%%%%%%%%%%%
\begin{abstract}

The problem of nucleon resonances $N'$ with masses below the 
$\Delta$ is considered.  We derive bounds for the
properties of such states. Some of these are new, while 
others improve upon existing limits. We discuss
the nature of $N'$ states, and their unitary partners, 
assuming their existence can be verified.  

\end{abstract}


\maketitle

%%%%%%%%%%%%%%%%%%%%%%%%%%%%%%%%%%%%%%%%%%%%%%%%%%%%%%%%%%
%%%   I. Introduction
%%%%%%%%%%%%%%%%%%%%%%%%%%%%%%%%%%%%%%%%%%%%%%%%%%%%%%%%%% 
\section{Introduction}
\label{sec:intro}

Baryon spectroscopy continues to motivate an extensive experimental
program, with most studies focused on the missing resonance problem.
While many states predicted by conventional quark models have yet
to be seen, other states, such as pentaquarks and hybrids, are also
interesting, as they offer potentially new information on the
dynamics of confinement. Given the underpopulation of conventional
3-quark states, it is difficult to identify these unconventional
states. If, however, a state was to be found with a mass between
the nucleon and $\Delta$, it would undoubtedly have an exotic
structure. 

Such a baryon state (called here $N'$, for brevity and according to 
tradition, though its isospin could be 3/2) was first suggested~
\cite{ya} to complete the unitary multiplet of hyperon resonance 
states $\Sigma(1480)$ and $\Xi(1620)$, considered now to have 
one-star status (see PDG listings~\cite{PDG02}).  A baryon state in 
the same mass interval was later suggested as a (quasi)bound 
pion-nucleon state (see sources in Ref.~\cite{ital}).  It appeared 
possible, even before any specially designed experiments, to obtain 
bounds for the properties of such a light baryon. Those bounds 
implied~\cite{ya,ital} that hadronic, and perhaps electromagnetic, 
couplings of the $N'$ to usual hadrons should be small (though not 
necessarily forbidden), thus suggesting a narrow resonance with a 
small production cross-section.  Missing mass experiments, as well 
as $\gamma N$ interactions and electroproduction, were suggested~
\cite{ya} as means to search for $N'$ states. 

Direct experimental searches for $N'$ have begun rather recently. 
Unfortunately, the results have been contradictory.  Initially,
in the reaction $pp\to nX^{++}$ at TRIUMF~\cite{ram} no baryon was 
detected with $I=3/2, m_N\leq m_X\leq m_N+m_{\pi}$ and a production
cross-section $>10^{-7}$ of the backward elastic $np$ cross-section 
(an additional assumption of a long lifetime was used).  However, 
in the reaction $pp\to p\pi^+X^0$ measured at Saclay~\cite{tat}
several low-mass structures were reported and interpreted as narrow 
peaks corresponding to new baryons.  

This report renewed interest, both theoretical and experimental,
in the subject.  If correct, such baryons would have isospin 
$I=1/2$, masses of 1004, 1044, and 1094~MeV, and widths less than 
4--15~MeV. Two of these could decay only radiatively, while for 
the third (slightly above the $\pi N$-threshold) the radiative 
decay channel could also be important.  The existence of these 
states was opposed in~\cite{L'W}, on the basis of their 
non-observation in the Compton scattering on protons or neutrons 
loosely bound in deuterons. 

Similar measurements of $pd\to ppX$ at INR (Moscow) gave evidence 
for structures~\cite{fil1} interpreted by the authors as 
corresponding to light narrow dibaryons (see~\cite{fil1,fil2} and 
references therein). Simultaneously, narrow structures with $B=1$ 
were also observed. These could be kinematically related to the 
dibaryons or correspond to new narrow baryonic states with masses 966, 
986, and 1003~MeV~\cite{fil1,fil2}  (the latter state perhaps related
to the 1004~MeV structure of~\cite{tat}). However, an attempt to study 
one of these reported dibaryons at RCNP, Osaka, in the same reaction, 
but with stated better mass resolution and better background conditions, 
showed no statistically significant effect~\cite{tam}, thus possibly 
casting doubt on both the narrow dibaryons and baryons of~\cite{fil1}. 

Narrow light baryons have been also searched for with good precision 
at JLab (Hall~A) and MAMI in electroproduction reactions $p(e,e'\pi^+)X$~
\cite{JLAB,MAMI} and $d(e,e'p)X$~\cite{MAMI}. No signals were found 
up to a missing mass of about 1100~MeV at the level ~$10^{-4}$ with 
respect to the height of the neutron peak.

The theoretical status of $N'$ resonances is similarly unclear. It 
was noted from the beginning~\cite{ya} that the smallness of 
$N'$-couplings to usual hadrons ``might be a consequence of the sharp 
difference in inner quark structure of $N$ and $N'$~".  Since the 
internal spin-flavor wave function for usual octet and decuplet 
baryons is totally symmetric, it has been assumed that new narrow 
baryons have a totally antisymmetric spin-flavor wave function~
\cite{kob}.  If so, they should not only have suppressed hadron 
couplings, but also forbidden one-photon decays.  Such a possibility 
looks attractive and is frequently referred to, since it could 
reconcile hadron production of $N'$ states with the absence of $N'$ 
signals in Compton scattering and electroproduction.  However, ground 
states (having $S$-wave space structure) with such spin-flavor 
properties should be, due to the Pauli principle, totally symmetric 
in color and therefore not colorless.

One explanation of the $N'$ states hypothesized~\cite{wal} the 
existence of a new ``light pion" with a mass of about 20~MeV.  New 
baryons were then assumed to be bound states of a usual nucleon with 
several ``light pions". However, existence of such ``light pions" has 
not been confirmed in any way.  Another suggestion~\cite{kon} has 
been to construct new baryons from clusters of diquarks.  The 
suggested mass formula produces a dense spectrum, able to accomodate 
all the reported states and many more.  Such approaches lie outside 
the mainstream of hadron physics, and are aimed mainly at a 
description of the reported mass spectrum of the narrow baryons. 

Our renewed investigation of the $N'$ puzzle has been partly
motivated by a recent set of measurements, suggesting that
unconventional multi-quark systems may indeed exist in nature.
Experimental evidence from SPring-8, ITEP, and JLab measurements~
\cite{nak,barm,kub,hic} suggests the existence of an exotic 
$\Theta^+$-baryon (former $Z^+$). Predicted~\cite{dpp} on the 
basis of the chiral soliton model, it has positive strangeness and, 
therefore, is exotic, \textit{i.e.}, cannot consist of only three 
quarks. If exotic hadrons really do exist, some could have the same 
quantum numbers as nucleons. The chiral soliton approach for 
$\Theta^+$ and its relatives (members of the same $SU(3)_F$-multiplet) 
predicts they will have $J^P=1/2^+$,  which requires, for the $(4q)\bar 
q$-system, at least one orbital excitation ($P$-wave). Therefore, one 
may expect the existence of lower-lying nucleon and other baryon 
states. We will return to this suggestion later on.

Our presentation proceeds as follows. In Section~\ref{sec:bound}, we 
first consider various new restrictions for the existence of $N'$ 
states, separately below and above the $\pi N$-threshold, and discuss 
how they are related.  Then, in Section~\ref{sec:nature}, we discuss 
the possibility of $N'$ being a candidate for a 5-quark system. 
We also give a tentative description of the unitary partners of $N'$. 
The whole picture is briefly summarized in the Conclusion.

\section{Bounds on $N'$ properties}
\label{sec:bound}

Having controversial results from dedicated experiments searching for 
the $N'$, we first study what limitations can be obtained at present 
from other considerations. This will allow us to check for consistency
in the present status of possible light nucleon resonances. It is 
convenient, at this point, to consider separately the cases of $N'$ 
states above or below the $\pi N$-threshold.   

%%%%%%%%%%%%%%%%%%%%%%%%%%%%%%%%%%%%%%%%%%%%%%%%%
\subsection{Elastic resonances }
\label{ssec:el_res}

If we assume that the new state $N'$ exists above the elastic $\pi 
N$-threshold, but below the $\Delta(1232)$, it is then natural to 
expect that $N'$ decays only (or, at least, mainly) to $\pi N$.  In 
this case, one might expect a partial wave analysis to easily reveal 
the presence or absence of such a resonance. This is, however, not 
quite so. 

There are two kinds of partial-wave analyses (PWA): single-energy 
(SE), when a PWA is made independently in narrow energy bins; and 
energy-dependent (ED), which uses an energy-dependent parametrization 
to consider simultaneously data at various energies. In the SE 
treatment, one can miss a resonance which is narrow enough to fall 
into the gap between two neighboring energy bins. The ED consideration 
assumes a mild energy dependence, and may smear a narrow resonance 
peak down to (nearly) zero. Consequently, we must use another approach 
to search for narrow elastic resonances.
  
We have used the $\pi N$ SAID database, which is the basis for SE and 
ED PWA's~\cite{SAID}.  The existence of a resonance was then assumed 
in a particular partial-wave amplitude (\textit{i.e.}, with fixed 
quantum numbers), having fixed values of mass and width. With this
addition, we have readjusted all other fitting parameters to minimize 
$\chi^2$. If a resonance is actually present, we expect that the fit 
should improve (lowering $\chi^2$) once it is included. 

We applied this procedure for pion laboratory energies below 500~MeV,
adding resonances to all $S$-waves, all $P$-waves, and two $D$-waves: 
$S_{11}, S_{31}, P_{11}, P_{13}, P_{31}, P_{33}, D_{13}$, and $D_{15}$. 
Other partial-wave amplitudes are very small in the considered energy 
interval and can be neglected. For trial masses, we use values from 
1100~MeV up to 1300~MeV (formally, we enter the inelastic region, but 
the inelasticity is very small).  For widths, we take 50, 100, 150, 
200, 250, and 300~keV (additional resonances with higher widths are 
definitely excluded).

\begin{figure}[th]
\centerline{
\includegraphics[height=0.5\textwidth, angle=90]{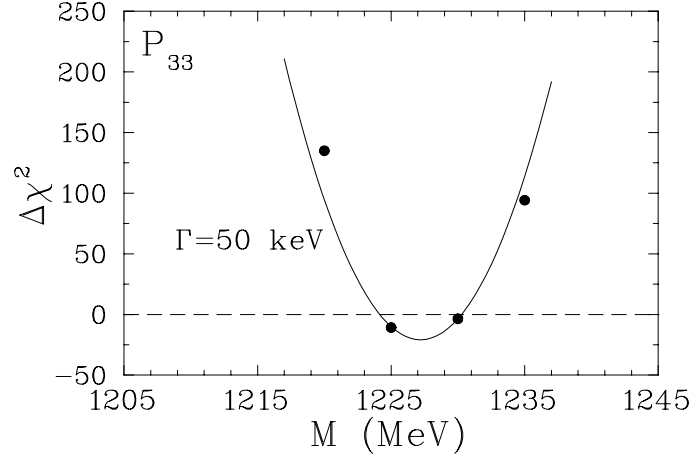}}
\caption{Change of overall $\chi^2$ due to insertion of a 
         resonance into $P_{33}$ for M = 1100 -- 1295~MeV 
         and $\Gamma$ = 50~keV, using $\pi N$ PWA~
         \protect\cite{pip_g2}. The curve is given to guide 
         the eye. \label{fig:g1}}
\end{figure}

\begin{figure}[th]
\centerline{
\includegraphics[height=0.5\textwidth, angle=90]{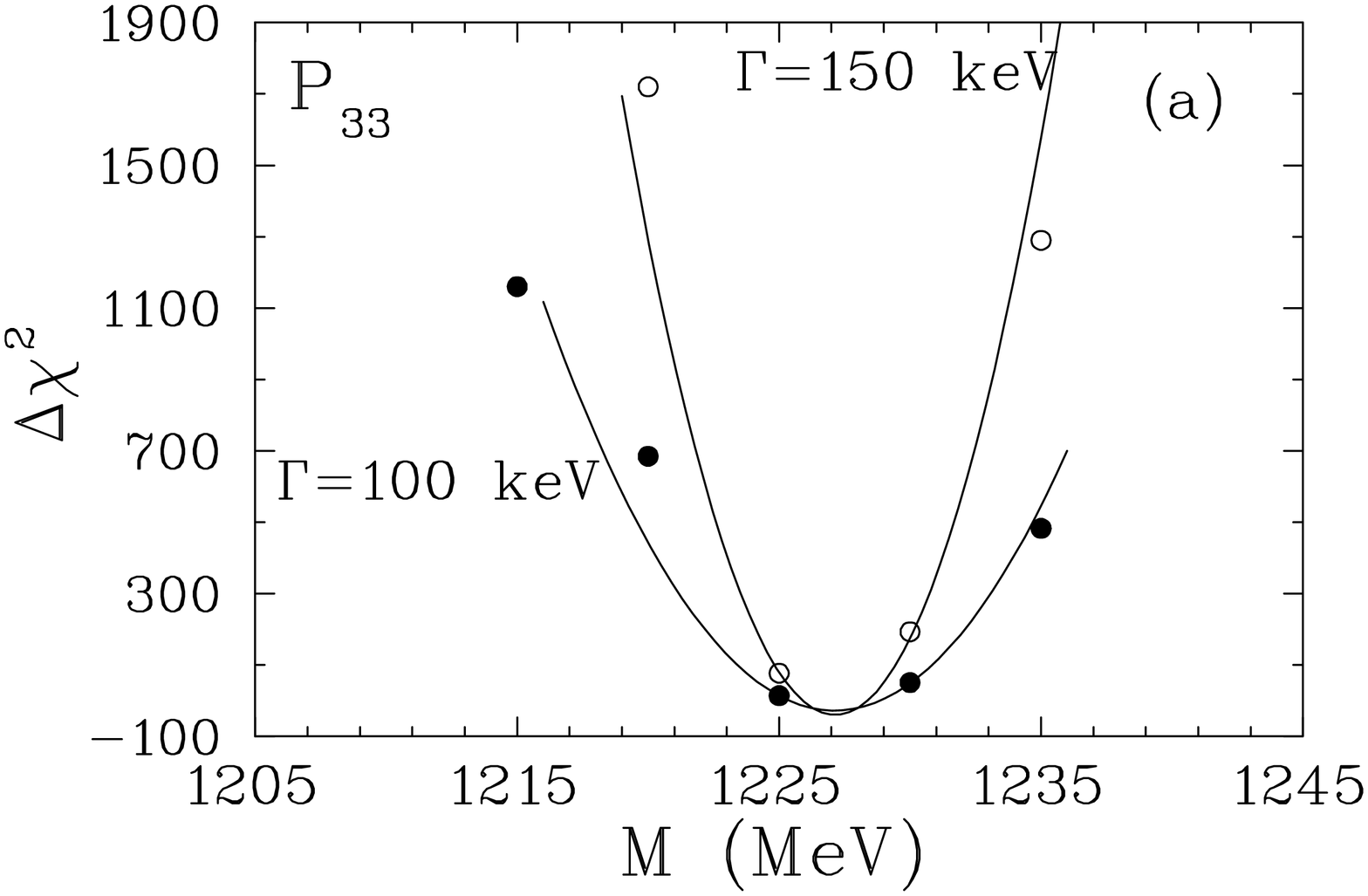}\hfill
\includegraphics[height=0.5\textwidth, angle=90]{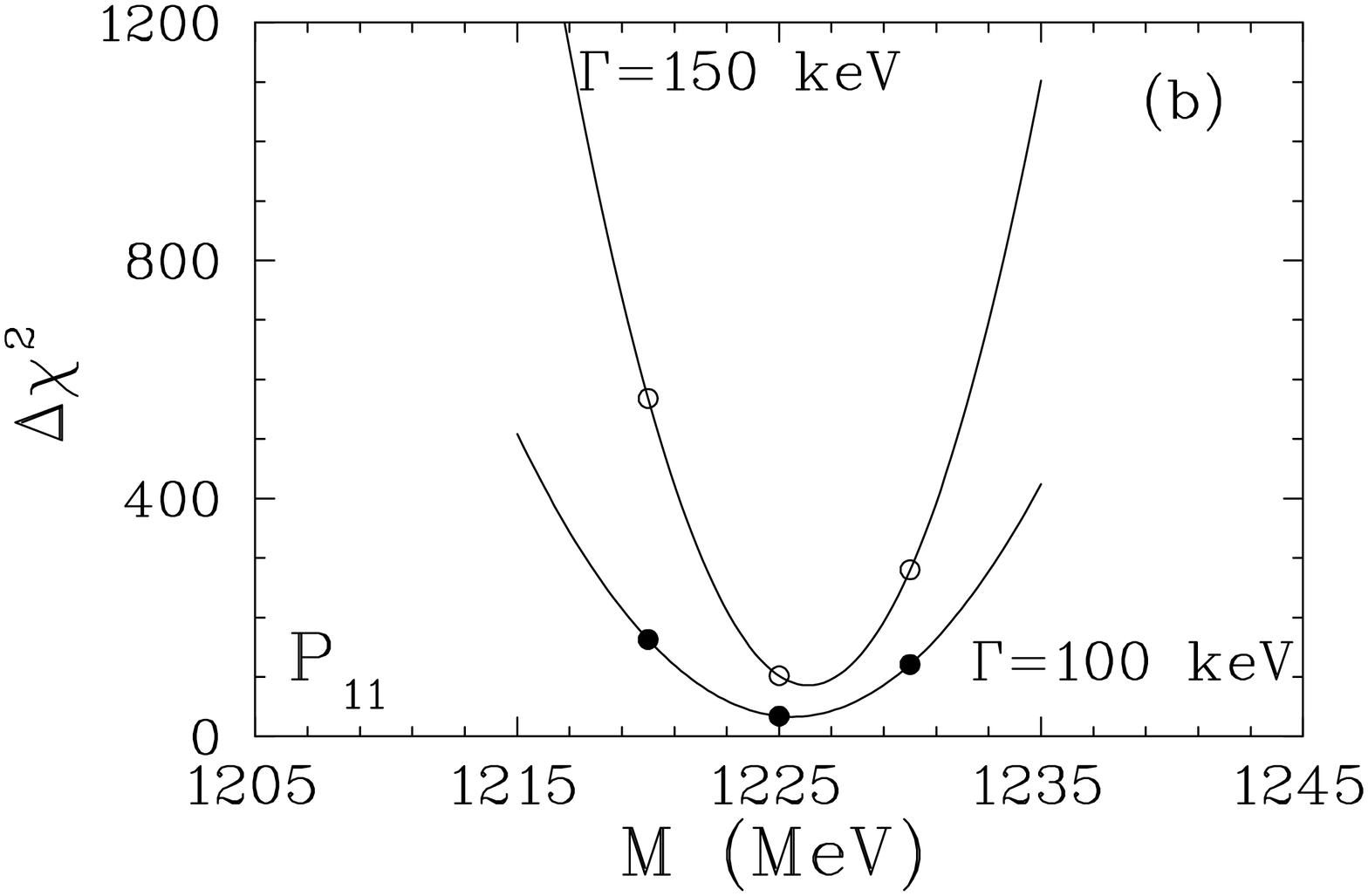}}
\caption{Change of overall $\chi^2$ due to insertion of a 
         resonance into (a) $P_{33}$ and (b) $P_{11}$ for 
         M = 1100 -- 1295~MeV and $\Gamma$ = 100 and 
         150~keV, using $\pi N$ PWA~\protect\cite{pip_g2}.
         The curves are given to guide the eye. \label{fig:g2}}
\end{figure}

Surveying our results, we found a case where it was possible to 
diminish $\chi^2$. This could be done by inserting a resonance with 
a mass of 1225~MeV and a width of 50~keV into the wave $P_{33}$ (see 
Fig.~\ref{fig:g1}). The change of $\chi^2$ reaches $-11$, while 
$\chi^2$ itself is about 6000.  To reveal the nature of this effect, 
we note that the ``suspected'' mass value appears very near the 
$\pi\pi N$-threshold which is 1220~MeV.  This threshold is accounted 
for in the parametrization of partial-wave amplitudes, but not exactly. 
Insertion of a narrow ``resonance'' imitates small corrections to the 
threshold description.  Such an interpretation is supported by the 
fact that $\Delta\chi^2$ as a function of the trial resonance mass has 
the local minima near 1220~MeV for any ``resonating" partial wave and 
for any assumed ``resonance" width (see, \textit{e.g.}, Fig.~
\ref{fig:g2}). 

\begin{figure}[th]
\centerline{
\includegraphics[height=0.5\textwidth, angle=90]{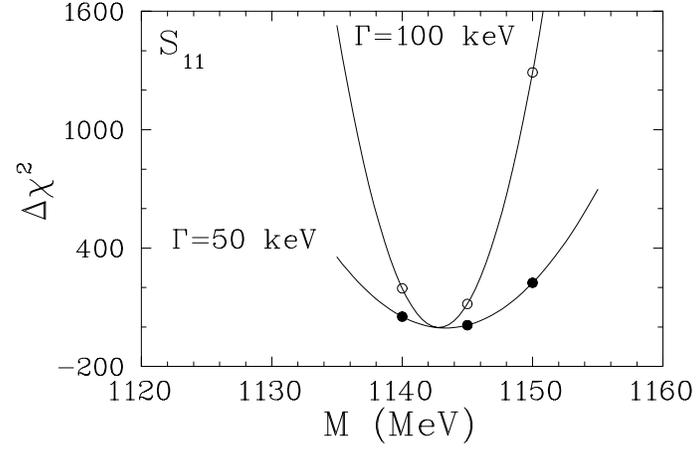}}
\caption{Change of overall $\chi^2$ due to insertion of a 
         resonance into $S_{11}$ for M = 1100 -- 1295~MeV 
         and $\Gamma$ = 50 and 100~keV, using $\pi N$ PWA
         ~\protect\cite{pip_g2}. The curves are given to 
         guide the eye. \label{fig:g3}}
\end{figure}

\begin{figure}[th]
\centerline{
\includegraphics[height=0.5\textwidth, angle=90]{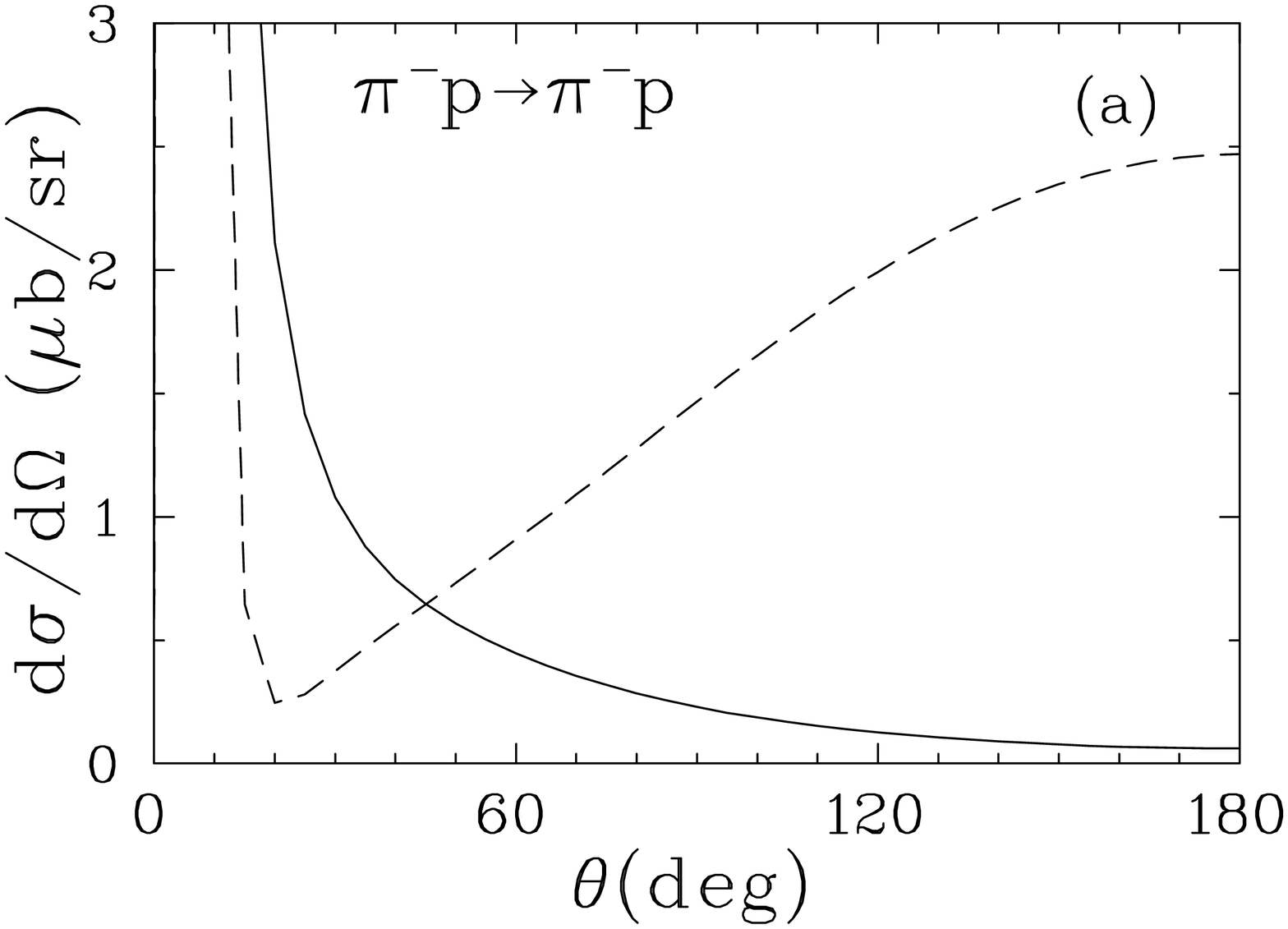}\hfill
\includegraphics[height=0.5\textwidth, angle=90]{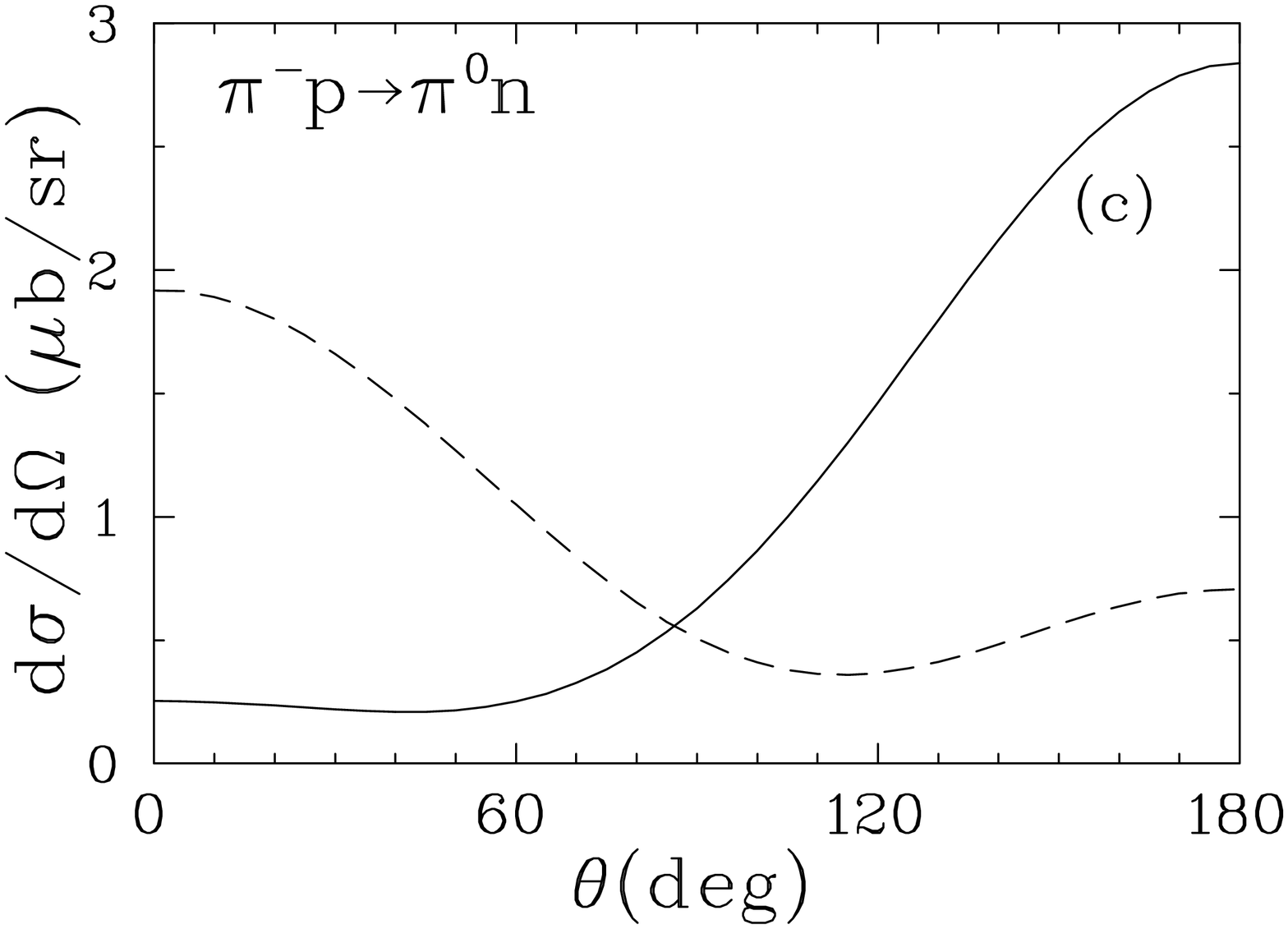}}
\centerline{
\includegraphics[height=0.5\textwidth, angle=90]{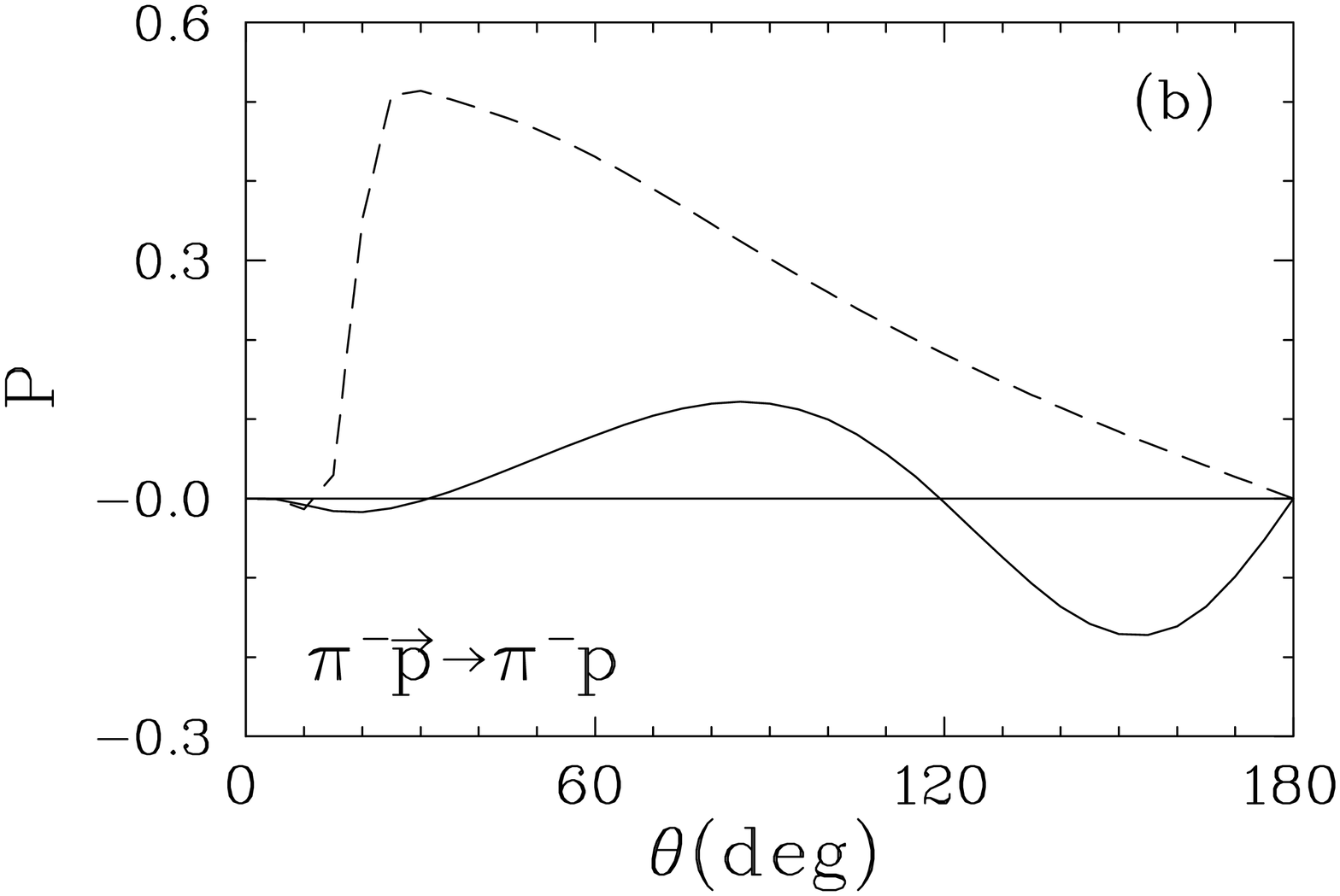}\hfill
\includegraphics[height=0.5\textwidth, angle=90]{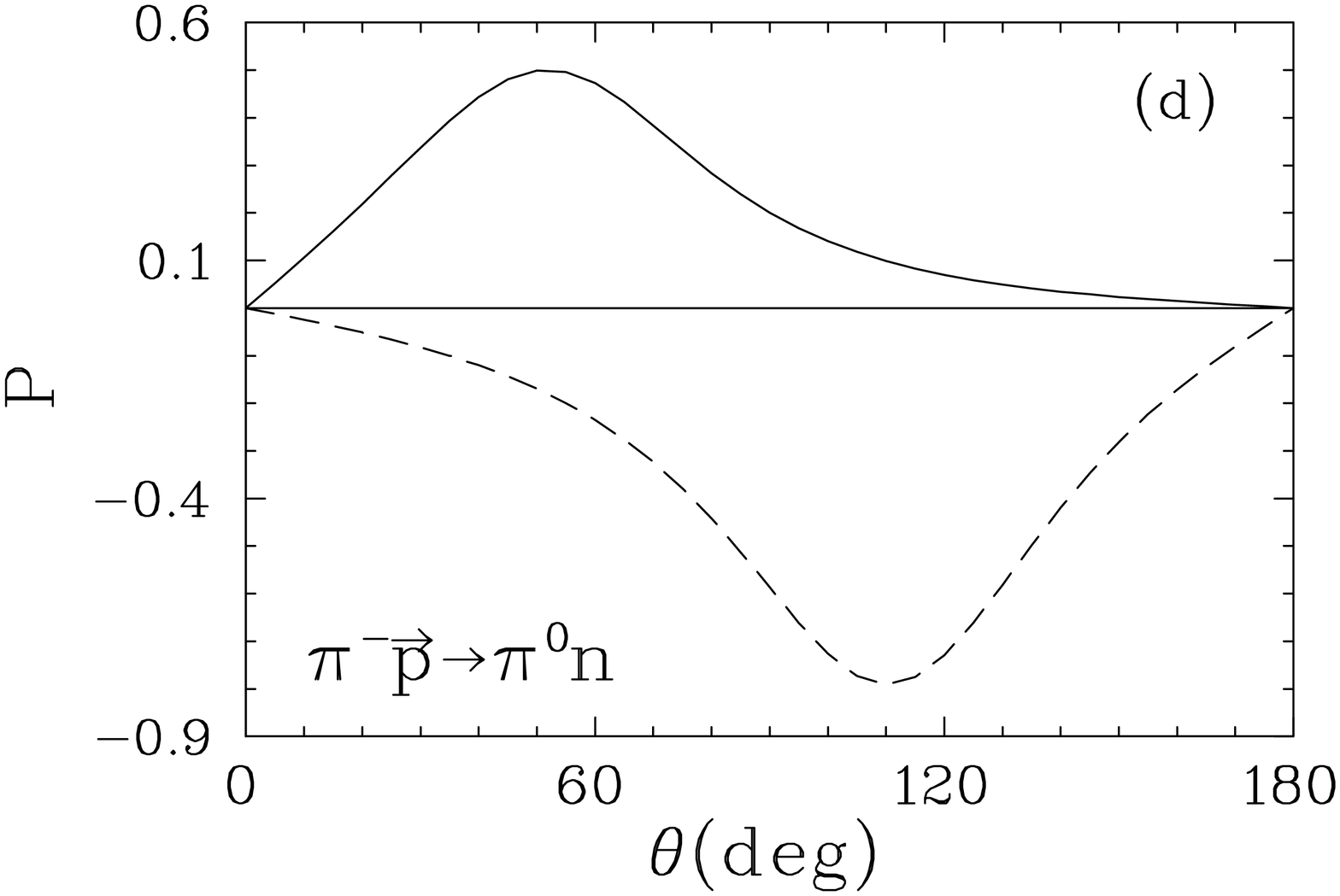}}
\caption{Differential cross sections (a,c) and polarization 
         parameter P (b,d) for $\pi^-p\to\pi^-p$ (a,b) and 
         $\pi^-p\to\pi^0n$ (c,d) at T$_{\pi}$ = 79.5~MeV.  
         The solid (dotted) line plots the SAID solution~
         \protect\cite{pip_g2} (plus the S$_{11}$ resonance
         at M = 1145~MeV and $\Gamma$ = 50~keV). \label{fig:g4}}
\end{figure}

\begin{figure}[th]
\centerline{
\includegraphics[height=0.5\textwidth, angle=90]{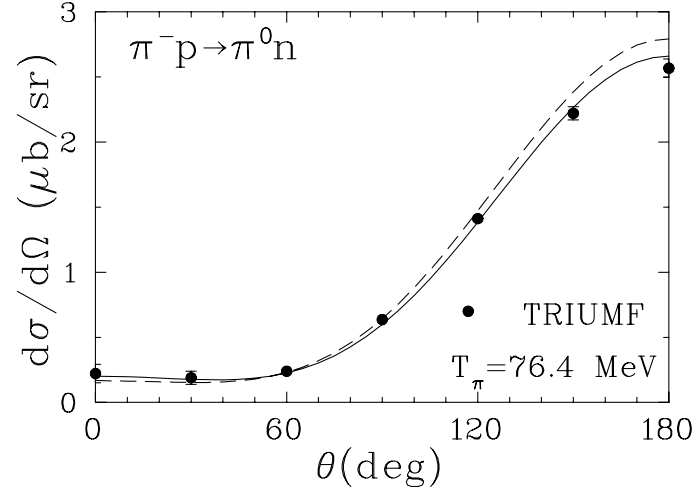}}
\caption{Differential cross section for $\pi^-p\to\pi^0n$ 
         at T$_{\pi}$ = 76.4~MeV.  The solid (dotted) line 
         plots the SAID solution ~\protect\cite{pip_g2}
         (plus the S$_{11}$ resonance at M = 1145~MeV and 
         $\Gamma$ = 50~keV). Experimental data at T$_{\pi}$ 
         = 76.4~MeV are from TRIUMF \protect\cite{cxs}. \label{fig:g5}}  
\end{figure}

One more interesting effect emerges in the wave $S_{11}$ for the 
resonance width $\Gamma=50$~keV. This generates a sharp minimum for 
$\Delta\chi^2$ at the assumed resonance mass 1145~MeV (which corresponds
to a pion kinetic energy of 79.5~MeV in the laboratory frame). Though
$\Delta\chi^2$ stays positive here, it takes a very small value, about 
9 (Fig.~\ref{fig:g3}). No threshold is present at this mass, and to 
clarify the case, we have examined the experimental data in this region. 
It appears that there is a gap in data, which could be ``filled'' by a 
narrow resonance (with a width smaller than 50~keV). Its presence 
would dramatically change cross-sections and polarization effects of 
$\pi N$ interactions in the resonance region as compared to the present 
non-resonant expectations (see Fig.~\ref{fig:g4}) but would have 
practically no effect on the existing data (Fig.~\ref{fig:g5}). 
Interestingly, this gap in data also allows local minima of 
$\Delta\chi^2$ near 1145~MeV for any partial wave and for each (small 
enough) trial resonance width. This situation demonstrates the 
limited sensitivity of existing data to the resonance problem. 
Indeed, sufficiently narrow resonances (with $\Gamma<50$~keV for
the present data) can always be inserted into one or another 
partial wave providing a better fit, even if a true resonance is 
absent there.

Our considerations allow us to draw some conclusions:

1) We find no evidence for elastic $\pi N$ resonances in the region 
between the $\pi N$-threshold and 1300~MeV having a width 
$\Gamma\geq50$~keV.

2) The present $\pi N$ data can not exclude even purely elastic 
(or inelastic) narrow resonances with widths below 50~keV. 

3) Insertion of trial narrow ``resonances" may be a good ``technical 
trick" to check the quality of fit to a set of experimental data. 

To estimate the meaning of the obtained results for additional 
resonance(s), let us compare them to the well-known properties of the
$\Delta(1232)$, having a width of about 120~MeV. Thus, we have
\begin{eqnarray}
\Gamma(N')< 50\,{\rm keV}\,,~~~~
\Gamma(N')/\Gamma(\Delta)< 4\cdot 10^{-4}.
\label{1}\end{eqnarray}

Up to now, we have discussed only the hadronic interactions of $N'$. 
However, such a narrow resonance could have a significant 
radiative decay $N'\to N\gamma$. If so, it should produce a signal 
in the Compton $\gamma N$ scattering, proportional to 
${\rm Br}^2_{\gamma}(N')\cdot\Gamma_{N'}$. Absence of the signal 
in the $\gamma p$ data up to $E_{\gamma}=290$~MeV~\cite{p-comp} 
allowed the derivation of a limit~\cite{L'W} which depends on the 
assumed mass of $N'$. For the whole region $m_N < m_{N'} < 
1200$~MeV, it gives
\begin{eqnarray}
{\rm Br}^2_{\gamma}(N')\cdot\Gamma_{N'} < 10\, {\rm eV}\,.
\label{2}\end{eqnarray}
For comparison, ${\rm Br}^2_{\gamma}(\Delta)\cdot
\Gamma_{\Delta}=3.6$~keV~\cite{PDG02}.
Thus, if the $N'$ does exist between the $\pi N$-threshold and the 
$\Delta$-region, the Compton data require a suppression
\begin{eqnarray}
\frac{{\rm Br}^2_{\gamma}(N')\cdot\Gamma_{N'}}
{{\rm Br}^2_{\gamma}(\Delta)\cdot\Gamma_{\Delta}} < 2.8\cdot 10^{-3},
\label{3}\end{eqnarray}
an order of magnitude weaker than the result of Eq.~(\ref{1})
for total widths.

%%%%%%%%%%%%%%%%%%%%%%%%%%%%%%%%%%%%%%%%%%%%%%%%
\subsection{Subthreshold states}
\label{ssec:sub-states}

We next consider $N'$ states below the $\pi N$-threshold. Of course, 
such states cannot decay to $\pi N$ and can not be seen as a resonance 
in $\pi N$ scattering. They may be, nevertheless, coupled to the $\pi 
N$ channel. Then, as was suggested earlier~\cite{ya,ital}, the $\pi N$ 
scattering data may give useful information about the $N'$ through  
dispersion relations (DR).
These relations for the $\pi^-p$ amplitude contain a contribution from
the neutron pole at the unphysical value $s=m_n^2$ ($s$ is the squared
$\pi N$ energy in the center-of-mass frame), with a residue proportional 
to $g^2_{\pi NN}$. The $\pi^+p$ amplitude does not contain such a pole, 
since there are no stable baryons with $I=3/2$, but has the neutron pole 
in the crossed channel, at the unphysical point $u=m_n^2$ ($u$ being the
squared 4-momentum transfer from proton to $\pi^+$, again, in the 
center-of-mass frame). These properties underlie the use of DR
to extract $g^2_{\pi NN}$ from experimental $\pi N$-scattering data 
(for a description of the procedure, see Ref.~\cite{AWP}). 

If the $N'$ does exist with $m_{N'}<m_N+m_{\pi}$ and couples to the 
$\pi N$ system, it generates an additional pole in the $\pi N$
scattering amplitude. For simplicity, let us assume here that $N'$ has
the same quantum numbers as the nucleon ($I=3/2$ is excluded with high
precision by the data~\cite{ram}; spin and/or parity of $N'$ different
from $N$ would only provide an additional factor, of order unity,
in the residue).
The procedure of~\cite{AWP} for such a case is really
sensitive only to the sum $g^2_{\pi NN}+g^2_{\pi NN'}$, and can not 
separate the two terms. Therefore, we should rewrite the result 
based on the use of DR~\cite{pip_g2} as
\begin{eqnarray}
(g^2_{\pi NN}+g^2_{\pi NN'})/(4\pi)=13.71\pm0.07. 
\label{4}\end{eqnarray}

There is, however, an alternative way to extract $g^2_{\pi NN}$ 
from the pion exchange contribution to $NN$ scattering. This is not 
spoiled by the presence of $N'$. A consistency requirement 
of the two methods 
can help to extract or restrict $g^2_{\pi NN'}$. In this way, 
$np$-scattering gives~\cite{np_g2}:  
\begin{eqnarray}
g^2_{\pi NN}/(4\pi)=13.69\pm0.09.
\label{5}\end{eqnarray}
Thus, $g^2_{\pi NN'}/(4\pi)$ should not be more than, say, 0.16, 
\textit{i.e.},
\begin{eqnarray}
g^2_{\pi NN'}/g^2_{\pi NN}\leq 10^{-2}\  .
\label{6}\end{eqnarray}
Note that an earlier bound of this kind was weaker, with a limit of 
0.1~\cite{ya,ital}. One should note, however, that the uncertainty
in Eq.~(\ref{5}) could be larger~\cite{np_study}.

A somewhat different method to restrict $g^2_{\pi NN'}$ was suggested 
in~\cite{ital}.  This was based on the Adler-Weisberger (AW) sum rule~
\cite{weis,adl} related to the algebra of currents. In contrast with
the DR method, the employed 
current algebra is not rigorously 
derived for strong interactions. It can only be an approximation 
requiring, in particular, the pion to be massless, without a 
systematic method for corrections.
Specifically for the AW sum rule, Adler has
discussed possible corrections, estimating the likely error to be
about 5\%~\cite{adl}. Therefore, methods based on the AW sum 
rule cannot give more reliable bounds than DR, and we do not use 
them here.   

As in the preceding subsection, we continue by considering 
processes including 
other interactions, which could be useful in the search for 
$N'$ states.  One of these is the capture of stopped pions.

Negative pions, being stopped in hydrogen, produce mainly two 
final states:
\begin{eqnarray}
\pi^-p\to n\pi^0\,,~~~~\pi^-p\to n\gamma\,. 
\label{7}\end{eqnarray}
Their relative abundance is described by the Panofsky ratio
$R=W(n\pi^0)/W(n\gamma)$ which is about 1.5. 

\begin{figure}[th]
\centerline{
\includegraphics[height=0.6\textwidth]{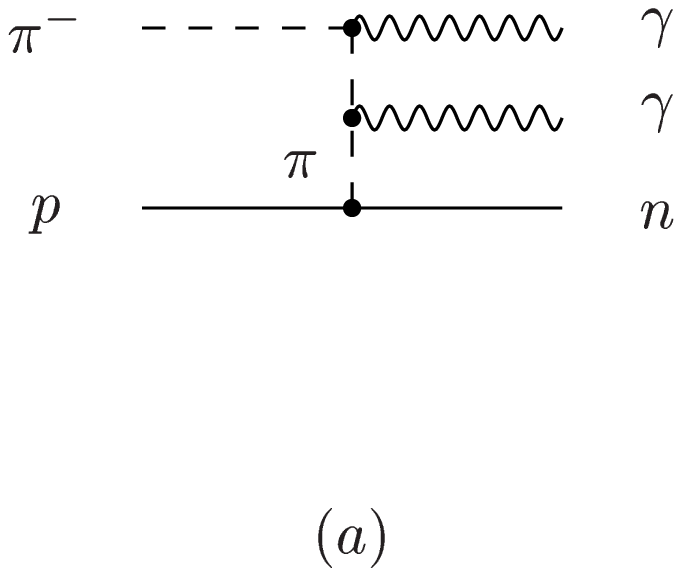}\hfill
\includegraphics[height=0.6\textwidth]{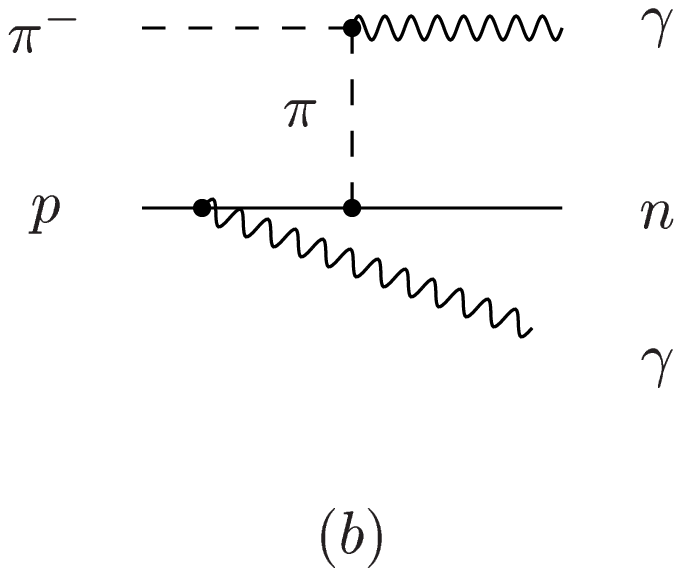}\hfill
\includegraphics[height=0.6\textwidth]{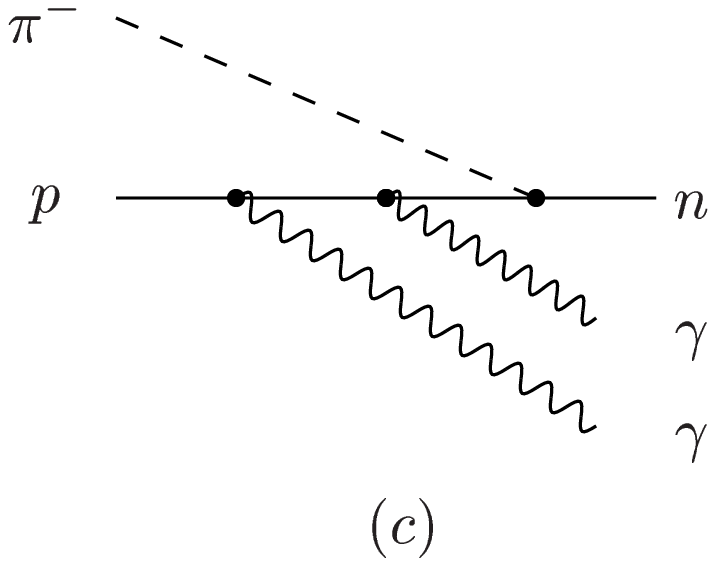}}
\caption{Diagrams for the direct $n2\gamma$ production in 
         $\pi^-p$-capture.\label{fig:g6}}
\end{figure}

The pion final state provides the best system for a precise determination 
of the pion mass difference, through accurate measurements 
of either the neutron velocity or that of the neutral pion. The 
former can be measured by the time of flight (TOF), while the latter 
can be found from $\gamma\gamma$ angular correlations in the final 
state $n2\gamma$ that emerges after the $\pi^0$-decay. The same final 
state, but with a different angular distribution, appears due to the 
direct two-photon process $\pi^-p\to n2\gamma$ (see diagrams of 
Fig.~\ref{fig:g6}).

An $N'$ state with mass between $m_n$ and $m_p+m_{\pi^-}$ provides 
one more source for the $n2\gamma$ final state:
\begin{eqnarray}
\pi^-p\to n'\gamma\to n\gamma\gamma
\label{8}\end{eqnarray}
(compare diagrams of Fig.~\ref{fig:g7}). Therefore,  a detailed 
investigation of this $2\gamma$ final state may provide further
evidence for, or a restriction on, $N'$ contributions.

The most precise measurement of the pion mass difference comes today 
from the TOF experiment at PSI~\cite{TOF}, with a nearly discrete 
neutron velocity corresponding to the $n\pi^0$ final state. One more 
discrete neutron velocity, for the $n\gamma$ final state, is also 
seen quite well. The direct transition to $n2\gamma$ and/or the 
$n'$-cascade would produce signals with different properties: they 
should have continuous velocity distributions. Unfortunately, such 
signals in the work~\cite{TOF}, if they exist, seem to be subtracted 
together with background.

Another approach was used in a TRIUMF measurement~\cite{2gam}. 
The authors have
studied the final $n2\gamma$ system in the kinematical configuration 
which totally excluded contributions from the $n\pi^0$ final state. They 
were thus able to find the signal for direct $n2\gamma$ decay. Assuming 
theoretically expected energy-angle distributions, the measured 
branching ratio for $\pi^-p\to n2\gamma$ was $[3.05\pm 0.27({\rm stat})\pm 
0.31({\rm syst})]\times10^{-5}$~\cite{2gam}.

Important for our goal here is the fact that the measured $\gamma\gamma$
distributions show reasonable agreement with theoretical calculations for
the direct $2\gamma$-decay. This means that 
up to statistical and systematic 
uncertainties (each about 10\%) there were no contributions of the
$n'$-cascade. Keeping in mind the incomplete kinematical coverage and
the
different energy-angle distributions for direct and cascade decays
(the latter depending also on the spin-parity of $n'$), we can safely use
the measured intensity of the direct decay as an upper bound for the
cascade decay. Then, accounting for the Panofsky ratio and assuming
a 100\%
branching ratio for $n'\to n\gamma$, we derive the conservative estimate
\begin{eqnarray}
\frac{W(\pi^-p\to n'\gamma)}{W(\pi^-p\to n\gamma)}< 8\cdot 10^{-5}~~[\sim 10^{-5}]\,. 
\label{9}\end{eqnarray}
The number in square parentheses corresponds to the assumption that 
contribution of the $N'$-cascade is smaller than the total experimental 
uncertainty of the direct decay signal. Again, note that earlier data 
on the $\pi^-$-capture allowed only a weaker result for this 
bound, $10^{-3}$~\cite{ya}. 

Coupling of $N'$ to the $N\gamma$ channel should generate a 
contribution to the Compton scattering. Since it has not been seen 
for proton or neutron targets, one obtains a mass-dependent 
bound for the radiative widths~\cite{L'W}. For the whole interval from 
$m_N$ up to the $\pi{N}$-threshold, it is 
\begin{eqnarray}
\Gamma (N'\to N\gamma)<5\, {\rm eV}\,, 
\label{10}\end{eqnarray}
while at the lower end of the interval it can be a fraction of an eV. 
In terms of dipole moments and their effective lengths this leads 
to values which can be 3 orders of magnitude smaller than the size of 
the nucleon~\cite{L'W}. Of course, $\Gamma(N'\to N\gamma)$ in the
discussed mass region is just  the total width $\Gamma_{N'}$ if this 
decay mode is not suppressed somehow. If, however, the $N\gamma$ mode 
is essentially suppressed, it might become comparable to the $N\gamma\gamma$ 
mode.

%%%%%%%%%%%%%%%%%%%%%%%%%%%%%%%%%%%%%%%%%%%%%%%%%%%%%
\subsection{Interpretation of bounds for $N'$}
\label{ssec:discus}

Let us summarize and compare existing bounds for various quantities
describing interactions (or couplings)  of the $N'$ with familiar hadrons. 
Results of both the preceding subsections and other works are compiled 
in the Table~\ref{tbl1}. At first sight, they can not even be compared 
to each other, since they concern different kinds of interactions and 
processes. However,  all those bounds are 
interrelated, at least, ``parametrically". 

To begin with, we consider first the case of an $N'$ below the
$\pi N$-threshold. States with $I=3/2$ are strongly excluded here
by the TRIUMF experiment~\cite{ram}. Keep in mind, however, that 
this strong bound is applicable only if the double-charged member 
of the isotopic quartet is very long-lived, having $\tau\geq10^{-2}$~sec;
for shorter lifetimes it becomes weaker. The bound is about $10^{-6}$,
instead of $10^{-7}$, for $\tau\approx25$~nsec and rapidly weakens for 
smaller $\tau$.  

For $I=1/2$, the most strict limitation in the Table~\ref{tbl1}
seems to be the bound for $\pi^-p\to n'\gamma$ compared to 
$n\gamma$. It is in good correspondence with limits from Compton
scattering, also strict. Indeed, using a suppression factor, say,
$5\cdot 10^{-5}$ we can estimate an upper bound for the radiative width 
of $N'$ as a function of its mass:
\begin{eqnarray}
\Gamma_{\gamma}(N')< 5\cdot 10^{-5}\,\Gamma_{\gamma}(\Delta)\cdot
(\frac{m_{N'}-m_N}{m_{\Delta}-m_N})^3\,.
\label{11}\end{eqnarray}
For masses 1004, 1044, and 1094~MeV this gives respectively 0.3, 1.5, 
and 4~eV as upper bounds of $\Gamma_{\gamma}(N')$, while direct 
treatment~\cite{L'W} of the Compton data provides in the same cases 
0.2, 1.6, and 7~eV.

\begin{figure}[th]
\centerline{
\includegraphics[height=0.8\textwidth]{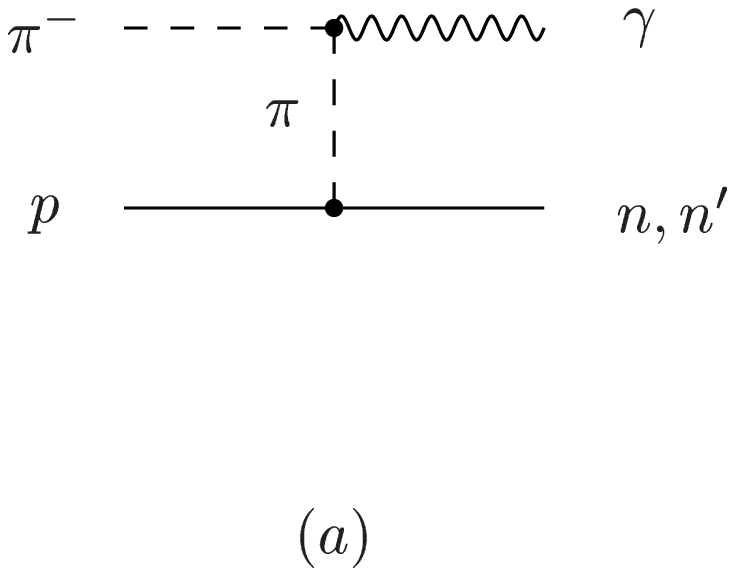}\hfill
\includegraphics[height=0.8\textwidth]{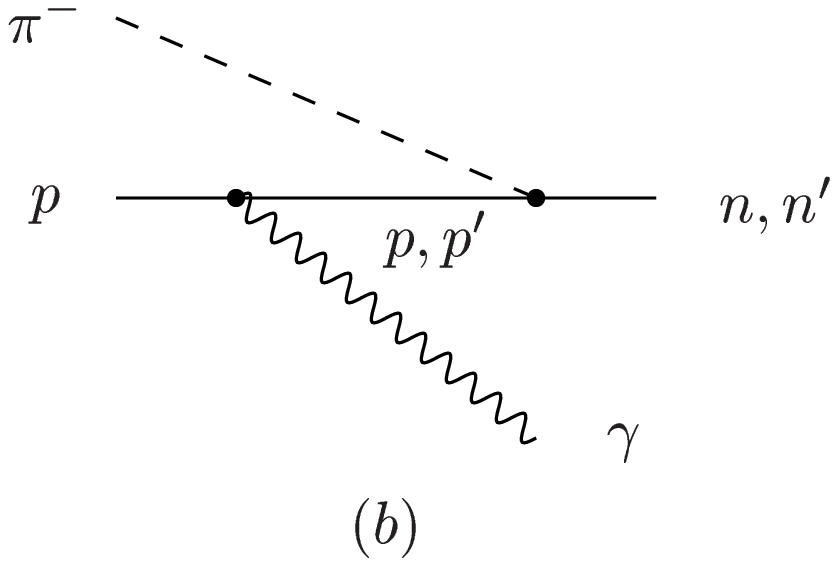}}
\caption{Diagrams for the radiative capture of $\pi^-p$. 
         \label{fig:g7}}
\end{figure}

At first sight, one can not directly compare estimates from 
$\pi^-$-capture to the bound for $g_{\pi NN'}^2$ from DR, because 
they relate to
different kinds of interactions. However, let us consider the structure 
of the corresponding amplitudes. Contributions to the radiative capture 
of the pion come from the diagrams like those of Fig.~\ref{fig:g7}. 
The main ones are the pion exchange (Fig.~\ref{fig:g7}a), proportional 
to $g_{\pi NN'}$ or $g_{\pi NN}$ in the amplitudes and to $g^2_{\pi NN'}$ 
or $g^2_{\pi NN}$ in the capture probabilities.  This illustrates that, 
if the capture to $N'\gamma$ has no special kinematical suppression 
(\textit{e.g.}, for $m_{N'}$ very close to $m_{\pi^-}+m_p$), then the 
radiative pion-capture limitation of Eq.~(\ref{9}) requires a stronger
suppression of the purely hadronic $N'$ couplings.  In particular,
\begin{eqnarray}
\frac{g^2_{\pi NN'}}{g^2_{\pi NN}}< 10^{-4}~~[10^{-5}]\, ,
\label{12}\end{eqnarray}
where the number in the square parentheses has the same meaning as in 
Eq.~(\ref{9}). Note that DR could provide only the weaker bound of 
Eq.~(\ref{6}), because of insufficient precision of the set of $\pi N$
and $NN$ scattering data.

Thus, $g_{\pi NN'}$ should be not more than $10^{-2}\cdot g_{\pi NN}$.
The presence of the non-pion-exchange contribution of Fig.~\ref{fig:g7}b,
without strong vertex supression, requires the radiative vertex 
$\gamma NN'$ to also be suppressed, in comparison with $\gamma NN$, at 
least by the same factor $10^{-2}$.  Moreover, Compton data show that in 
some cases the radiative vertex may be suppressed even stronger by the
factor of $10^{-3}$~\cite{L'W}.  The situation for the case of 
$m_{N'}>m_N+m_{\pi}$ looks similar. 

We can make the self-consistent assumption that in all cases both 
strong and electromagnetic couplings of $N'$ with usual hadrons 
should be suppressed more strongly than by a factor of $10^{-2}$ in 
amplitudes.

As a result, we expect that if the light resonances do exist, 
their hadro-, photo- and electroproduction can be seen only at a 
level smaller than $10^{-4}$ with respect to ``normal" cross-sections 
for usual hadrons. We also note in passing 
that for $m_{N'}< m_N+m_{\pi}$ the 
hadroproduction of $N'$ could appear as a special contribution to 
bremsstrahlung, \textit{e.g.}, $NN\to NN'\to NN\gamma$. 

%%%%%%%%%%%%%%%%%%%%%%%%%%%%%%%%%%%%%%%%%%%%%%%%%%
\section{Possible nature of $N'$}
\label{sec:nature}

The bounds for $N'$ properties, discussed above, appear 
rather severe and may be considered as evidence against 
the existence of such states. If, nevertheless, there are 
arguments for their existence, one needs to have an 
explanation for why couplings to usual hadrons are so 
suppressed.

In the Introduction, we briefly mentioned a motivation for
considering non-standard quark states, based on the 
recently reported baryon $\Theta^+$~\cite{nak,barm,hic,kub} with 
clearly exotic quantum numbers.  Being identified on the basis 
of rather low statistics, further confirmation is necessary.
However, if it does exist, it poses questions for hadrons with 
non-exotic quantum numbers as well. Here, we discuss the 
$\Theta^+$ and its possible relation to the $N'$ problem
in more detail.

$\Theta^+$ has strangeness $S=+1$ and, being considered as a quark 
system, should contain at least 4 nonstrange quarks and 1 strange 
antiquark. Its experimental mass agrees very well with a theoretical 
prediction~\cite{dpp}. This gives some hope that its spin and 
parity also correspond to the predicted values $J^P=1/2^+$. However, 
the product of internal parities of 4 quarks and 1 antiquark is 
negative. Therefore, the space wave-function of $\Theta^+$ can
not be pure $S$-wave; it should contain at least one $P$-wave to
make the total parity be positive. 

In quantum theory (at least, non-relativistic) there exists
a mathematically exact result that the space wave function 
of the ground (lowest-energy) state should not have zeros.
Since the $P$-wave Schr\"odinger function inevitably has
at least one zero, the ground-state character of $\Theta^+$
may be questionable.  Of course relativistic theory has some 
specifics, and there are recent statements~\cite{ris} that in 
the particular case of the quark structure of $\Theta^+$, the 
hyperfine interaction may reverse the normal order of the lowest 
$S$- and $P$-wave states.  However, the flavor dependence 
of such interaction prevents this property from being universal 
for all members of the $SU(3)_F$ antidecuplet which contains 
$\Theta^+$. Therefore, if the non-strange partner of $\Theta^+$ 
is indeed $N(1710)$, as assumed by~\cite{dpp}, we can expect 
that it is not a ground state.

Dynamics of the 5-quark system may be rather unfamiliar.
Nevertheless, having nothing better at present, we can try to 
use 3-quark experience for a tentative estimate of the 
energetic ``price" of a $P$-wave in a system with nucleon 
flavor quantum numbers.  

The ground state for baryons with $S=0$ and $I=1/2$ is $N(940)$ 
with $J^P=1/2^+$. It corresponds to the 3-quark system having 
the pure $S$-wave space function and sum of the spins equal 
1/2. If we consider the corresponding excited system with 
one $P$-wave, we obtain two states with $J^P=1/2^-$ and $3/2^-$ 
having different masses due to $({\mathbf LS})$-coupling. 
Particle Tables~\cite{PDG02} show that the lowest states with such 
quantum numbers are $N(1520)$, with $J^P=3/2^-$, and $N(1535)$, 
with $J^P=1/2^-$, both having the highest 4-star status. We see, 
therefore, that the $({\mathbf LS})$-coupling is relatively weak,
while the $P$-wave excitation requires about 600~MeV. 

\begin{figure}[th]
\centerline{
\includegraphics[height=1.2\textwidth]{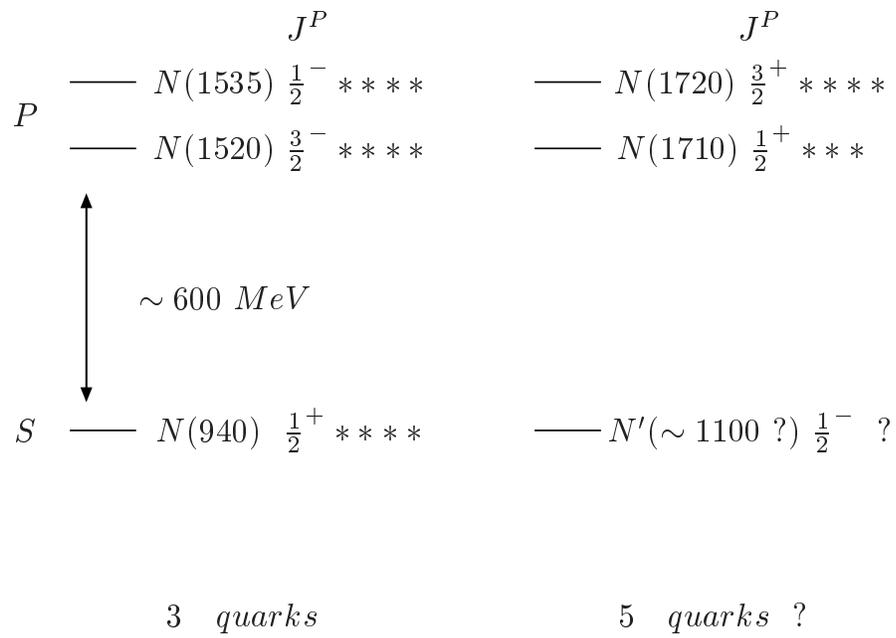}}
\caption{Possible S- and P-wave levels in quark systems. 
         \label{fig:g8}}
\end{figure}

Near $N(1710)$, with $J^P=1/2^+$ and 3-star status, we find
$N(1720)$, with $J^P=3/2^+$ and 4-star status. If they both 
are 5-quark systems with one $P$-wave, having the same
energetic ``price" of about 600~MeV, we expect that the 
corresponding ground state should have mass about 1100~MeV.
Thus, it is just the mass region near the $\pi N$-threshold where 
appearance of an $N'$ is expected. The situation is schematically 
shown at Fig.~\ref{fig:g8}. By analogy with usual hadrons, we show 
quantum numbers of $N'$ as $J^P=1/2^-$. However, the 5-quark system
is, of course, complicated enough, and may manifest several states 
with nearby masses, having different values of $J^P$. 

%%%%%%%%%%%%%%%%%%%%%%%%%%%%%%%%%%%%%%%%%%%%%%%%%%%%%%
\subsection{Problem of suppressed couplings}
\label{ssec:coupl}

So far the picture of an $N'$ as a 5-quark state looks
sufficiently consistent. But, as we explained above, to support 
it, we should demonstrate that such a picture has the ability to
describe the phenomenologically necessary suppression of 
couplings of $N'$ with usual 3-quark baryons. 

Dynamics of the 5-quark system may be essentially different 
from that of the 3-quark system.  Even the constituent-quark mass,
being a dynamical quantity, might be different for these two cases
(most probably, it decreases with increasing the quark number). 
That is why we will not pretend here to give a reliable 
description of coupling constants for the 5-quark hadrons. However, 
we can recall some known phenomena which may provide a realistic 
basis to describe the suppression of couplings.

At first sight, the 5-quark baryon can be easily separated into a
usual baryon (3 quarks) and a usual meson (quark and antiquark).
But this may be difficult because of inappropriate color 
structure. In this connection, let us recall the color suppression,
well known in weak decays (especially, of heavy-quark mesons).

\begin{figure}[th]
\centerline{
\includegraphics[height=0.6\textwidth]{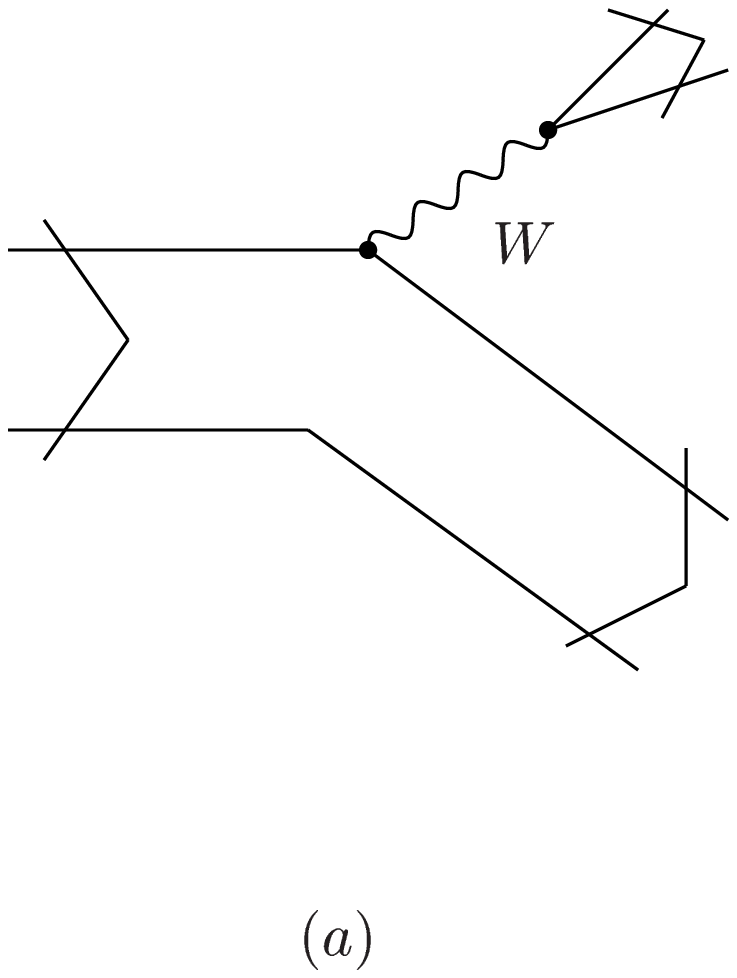}\hfill
\includegraphics[height=0.6\textwidth]{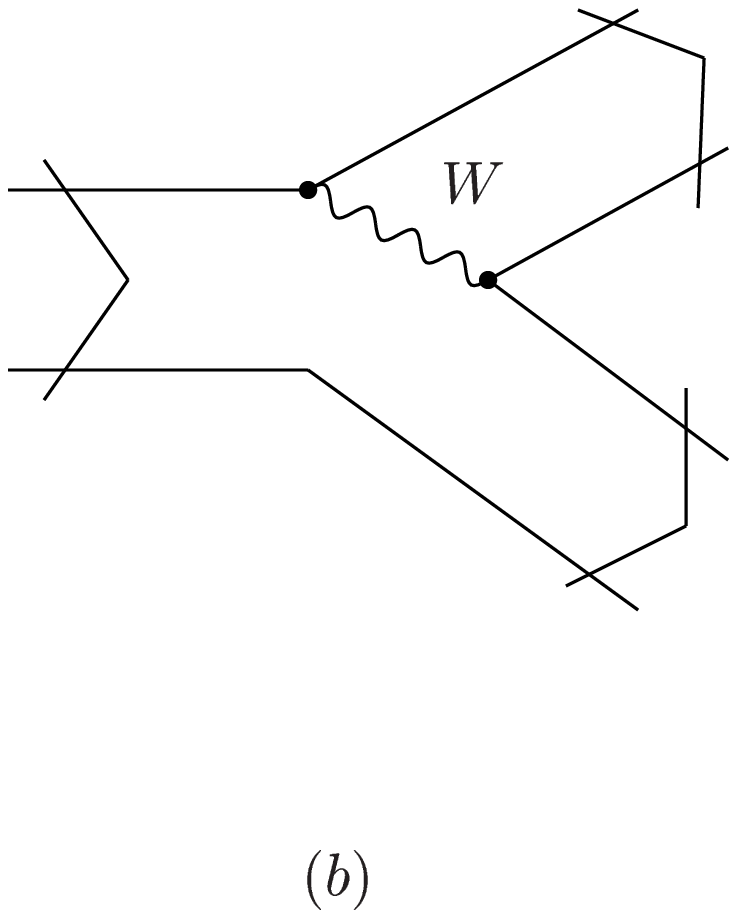}}
\caption{Decay diagrams without/with color supression. 
         \label{fig:g9}}
\end{figure}

Fig.~\ref{fig:g9} shows two kinds of contributions for weak decays. 
In both cases, the $W$-boson produces the colorless quark-antiquark 
pair. In one case (Fig.~\ref{fig:g9}a), the pair directly transforms 
into a meson (\textit{e.g.}, $\pi$-meson), without any problem. In 
the other case (Fig.~\ref{fig:g9}b), the quark and antiquark, 
separately, produce hadrons together with other quarks and 
antiquarks of the system. Not all color configurations of the pair 
are appropriate for the second process, so its amplitude contains the 
factor $1/N_c$, and its probability contains $1/N_c^2$, where $N_c$ is 
the number of colors. Thus, at $N_c=3$ such simple ``color 
suppression", even in decays of ``normal" hadrons, provides a
factor about 1/10 for the probability of the 
``suppressed" final state.  

The increased number of quarks in the system should increase
the number of possible inconsistencies in its color structures,
which suppress decays of the system. Double color suppression 
by itself would give the suppressing factor $10^{-1}$ for the 
strong coupling between a 5-quark baryon and, say, the 
baryon-meson pair of usual octet hadrons. Together with similar 
inconsistencies of the flavor and spin parts of the wave function,
it may be not so hard to provide a suppression of $10^{-2}$ for 
the coupling constants of $N'$, \textit{i.e.}, $10^{-4}$ for 
processes of its production.     

If the color and spin-flavor structures of the 5-quark baryon
are indeed capable of producing a suppression of $10^{-2}$ or 
more for strong couplings of the 5-quark baryon, they should
give, at least, the same suppression for the photon vertex of 
such a baryon. However, as we discussed in the preceding section,
the phenomenological photon vertex may need even stronger 
suppression, at least, $10^{-3}$. Let us consider whether this 
could be realistic.

\begin{figure}[th]
\centerline{
\includegraphics[height=0.8\textwidth]{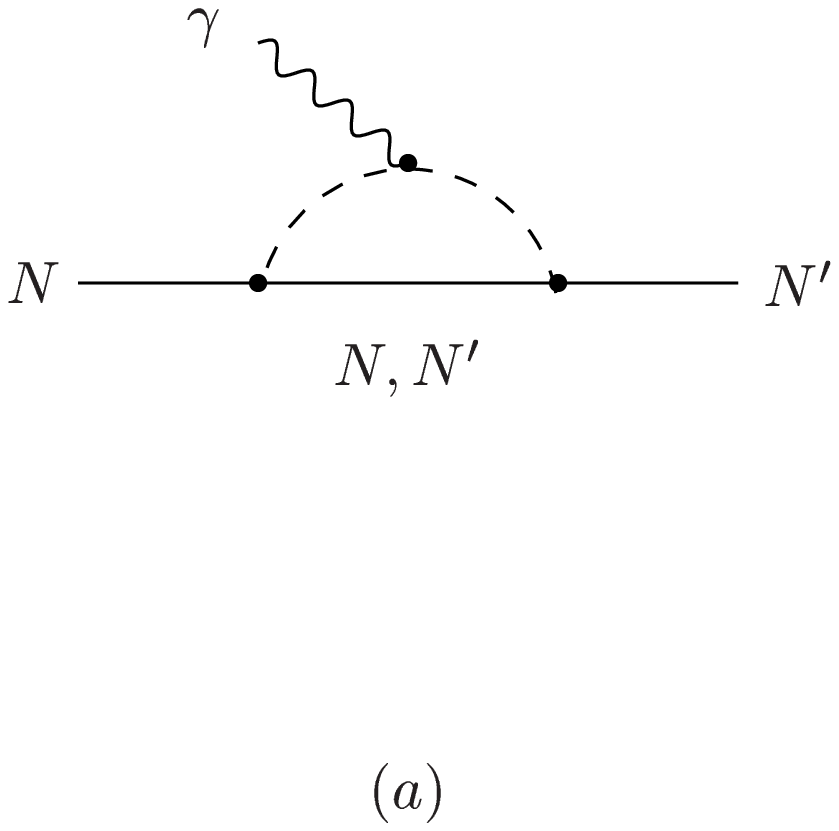}\hfill
\includegraphics[height=0.6\textwidth]{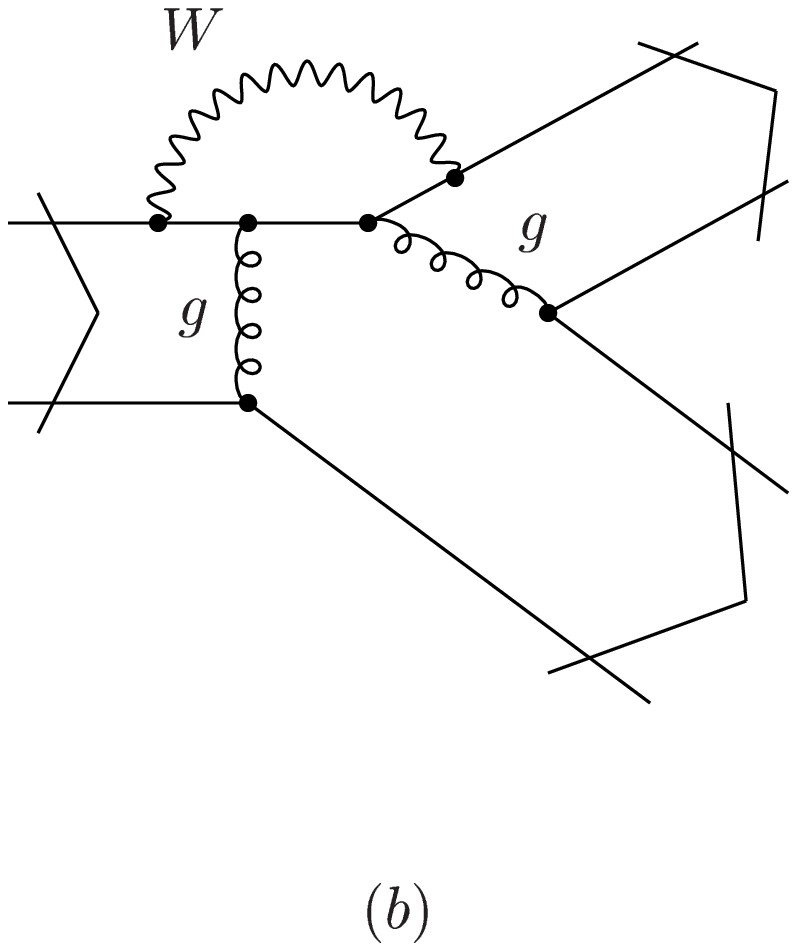}}
\caption{Loop diagram contributions to decay vertices.
         (a) Diagram for $N'N\gamma$.  (b) Penguin diagram 
         for weak decay.
 \label{fig:g10}}
\end{figure}

In the framework of the constituent quark model, the diagonal 
and transition dipole moments (say, magnetic moments) for 
usual (octet and decuplet) baryons can be well described 
as simple matrix elements of the single-quark electromagnetic 
interaction between quark wave functions of the initial and 
final baryons. But such a simple approach can not work for the 
photon transition between 5- and 3-quark baryons, because of the
different number of quarks. This vertex should have a more 
complicated structure, \textit{e.g.}, that of Fig.~\ref{fig:g10}a. 
It evidently contains the suppression of strong couplings, but 
its loop configuration may provide additional suppression,  
similar to the so-called ``penguin" diagrams of 
Fig.~\ref{fig:g10}b in weak processes. Those diagrams do not 
have parametric smallness with respect to usual weak amplitudes, 
but are known to be numerically small.  

The existence of diagrams like Fig.~\ref{fig:g10}a shows that the 
1-photon transition between $N'$ and $N$ may be suppressed, but 
can not be forbidden entirely, contrary to the suggestion of~
\cite{kob}. In such a situation, an interesting question arises
as to 
whether the suppressed probability of the 1-photon decay for 
$N'$ might become numerically of the same order as probability 
of the 2-photon decay.

%%%%%%%%%%%%%%%%%%%%%%%%%%%%%%%%%%%%%%%%%%%%%
\subsection{Unitary partners of $N'$}
\label{ssec:partners} 

With the existence of an $N'$  there inevitably emerge
additional problems,
related to the $SU(3)_F$ symmetry. What is its unitary 
multiplet?  And what are its unitary partners?

\begin{figure}[th]
\centerline{
\includegraphics[height=0.5\textwidth, angle=90]{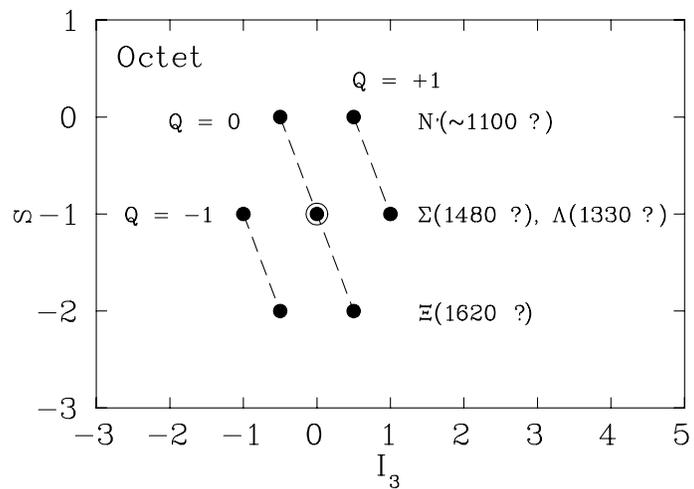}}
\caption{Tentative unitary octet with $N'$. \label{fig:g11}}
\end{figure}

Both questions require detailed investigation which will be 
given elsewhere. For now, we restrict ourselves to the 
simplest hypothesis of $N'$ being a member of a unitary octet, 
and tentatively discuss other possible members of this octet 
(Fig.~\ref{fig:g11}).

Two of the potential candidates appear to be present in the PDG 
listings~\cite{PDG02}. They are $\Sigma(1480)$ and $\Xi(1620)$, 
both with low 1-star status. One more multiplet member might be 
the resonance $\Lambda(1330)$ observed as a peak in the system 
$\Lambda\gamma$~\cite{bud}. 

All these states were observed in experiments with bubble 
chambers, and have been nearly forgotten with the coming of a new 
generation of detectors and facilities (and also new energy 
regions). The latest publications are~\cite{pan,eng} on 
$\Sigma(1480)$ and~\cite{ross,brief} on $\Xi(1620)$. 

But recently they have begun to reappear. $\Sigma(1480)$ is 
seen in very preliminary data of COSY~\cite{kop}, weak evidence 
for $\Lambda(1330)$ may be seen in a low statistics preliminary 
study of the $\Lambda\gamma$ spectrum at JLab (Hall B)~\cite{dis}. 
$\Xi(1620)$ has recently emerged in theoretical calculations of 
$\Xi\pi$-scattering in the framework of a unitary extension of 
chiral perturbation theory~\cite{ramos,oset}. Interestingly, 
these calculations assign $J^P=1/2^-$ for $\Xi(1620)$, exactly as 
we suggested above for $N'$.  Moreover, the Gell-Mann-Okubo mass 
formula with masses of $\Lambda(1330), \Sigma(1480)$, and 
$\Xi(1620)$ gives for $N'$ just the mass of about 1100~MeV~\cite{ya}, 
in agreement with the estimation above, based on different arguments. 
Present information on this tentative unitary octet is summarised 
in the Table~\ref{tbl2}. It shows, in particular, reported decay 
modes and values of hadronic production cross-sections. Note that 
the corresponding cross-sections for photoproduction may be 
estimated as multiplied by the factor $\alpha/\pi$, while for
electroproduction, the factor should be of order $(\alpha/\pi)^2$.

Of course, the experimental status of all these states is quite 
uncertain. Publications, which report their observation, estimate 
their statistical significance at the 3, or even 4, standard 
deviation level (for $\Sigma(1480)$ both the peak in the mass 
distribution and the polarization effect were reported~\cite{pan}). 
Many papers, which do not support those states, actually see the 
corresponding peaks, but can not exclude their non-resonant origin 
(background fluctuations, kinematical reflections and so on). 
Therefore, the problem should be further investigated at the modern 
level of accuracy. 

%%%%%%%%%%%%%%%%%%%%%%%%%%%%%%%%%%%%%%%%%%%%%%%%%%%%%
\section{Conclusion and discussion}
\label{sec:conclus}

The recent discovery of $\Theta^+$~\cite{nak,barm,hic,kub} (of 
course, being reliably confirmed) may open a new vista on the
field of many-quark hadrons. Their dynamics, though also based 
on QCD, can be phenomenologically different from the familiar 
strong interactions of the standard 3-quark and quark-antiquark 
hadrons.  Among other opening possibilities, there could (or 
even should) exist new light nonstrange baryon(s), with mass(es) 
near $N$ and $\Delta$. 

In this paper, we have studied the present bounds on properties
of the hypothetical light baryon(s) $N'$. Together with the 
dedicated experiments searching for $N'$, we also consider other 
data, not obviously directly related to $N'$. Using these, 
we are able to enhance previous bounds, and obtain new ones, 
for both strong and electromagnetic couplings of the $N'$. 
 
While $\Delta$-like baryons (with $I=3/2$) below the $\pi 
N$-threshold are strongly excluded at the level of $10^{-7}$~
\cite{ram}, it is not so for $N$-like states (with $I=1/2$) in the 
same mass region. Here, we show that all couplings of $N'$ to the 
standard hadrons should be suppressed more strongly than a factor 
of $10^{-2}$. This implies small (radiative) decay widths and 
small production cross-sections (less than $10^{-4}$ or even 
$10^{-5}$ with respect to analogous production of standard hadrons). 
Above the $\pi N$-threshold and up to the $\Delta$-region, we 
obtain new restrictions for couplings of both $I=1/2$ and 3/2 
nonstrange baryons, again at a level stronger than $10^{-4}$. 
Though the 5-quark systems and their dynamics are complicated and 
insufficiently understood, we give arguments that the necessary 
phenomenological suppression may be realistic. 

We have also briefly discussed unitary multiplets possibly related  
to $N'$ and 5-quark systems. They could be both familiar octets 
and decuplets, and also clearly exotic antidecuplets or even 
27-plet(s) (note that $\Delta$-like states do not appear in octets 
and/or antidecuplets). We have recalled some nearly forgotten 
states which could appear as unitary partners of $N'$. Studies of 
such partners might give an alternative view of the problem of 
$N'$. It is interesting in this connection that the reported 
cross-sections for hadronic production of those states (of order 
of several $\mu$b, see Table~\ref{tbl2}) are consistent with rough
estimates of several nb for photoproduction of $\Theta^+$~
\cite{poku} (the relative factor of $\alpha/\pi$).

The problem of $N'$ may have even broader interest than just 
hadron physics. For instance, it was demonstrated recently that 
existence of $N'$ may influence properties of neutron stars~
\cite{astro} and diminish their mass. Since this result was used 
by the authors as an argument against the existence of $N'$ (the 
calculated limiting mass of the neutron star appears lower than 
the experimental value), we would like to note that similar 
problems might arise also due to (well established) hyperons. 
They, however, may be eliminated by other effects, such as 
rotation excitations, repulsive potentials and other effects~
\cite{nstar}, which were not accounted for in~\cite{astro}.

%%%%%%%%%%%%%%%%%%%%%%%%%%%%%%%%%%%%%%%%%%%%%%%%%%%%%%%%%
%%%   Acknowledgments
%%%%%%%%%%%%%%%%%%%%%%%%%%%%%%%%%%%%%%%%%%%%%%%%%%%%%%%%%
\acknowledgments

One of the authors (Ya.~A.) highly appreciates the hospitality 
extended to him by the Jefferson Lab.  
The authors acknowledge useful communications with
W.J.~Briscoe, X.~Jiang, V.~Koubarovsky, M.~Kovash, E.~Pasyuk,
G.~Pavlov, M.~Polyakov, W.~Roberts, and B.~Wojtsekhowski.
This work was supported in part by the 
U.~S.~Department of Energy under Grant DE--FG02--99ER41110
and Russian State Grant SS--1124.2003.2.  
The authors acknowledge partial support from Jefferson Lab, 
by the Southeastern Universities Research Association under 
DOE contract DE--AC05--84ER40150.

\eject
%%%%%%%%%%%%%%%%%%%%%%%%%%%%%%%%%%%%%%%%%%%%%%%%%%%%%%%%%
%%%    V. References
%%%%%%%%%%%%%%%%%%%%%%%%%%%%%%%%%%%%%%%%%%%%%%%%%%%%%%%%%

\newpage
%%%%%%%%%%%%%%%%%%%%%%%%%%%%%%%%%%%%%%%%%%%%%%%%%%%%%%%%%
%%%  Tables
%%%%%%%%%%%%%%%%%%%%%%%%%%%%%%%%%%%%%%%%%%%%%%%%%%%%%%%%%
%%%%%%%%%%%%%%%%tbl.1
\begin{table}[h]
\caption{Bounds for $N'$ properties. \label{tbl1}}
\begin{center} 
\begin{tabular}{c|c|c}
\hline
Interactions & Below $\pi N$-threshold & Above $\pi N$-threshold \\
\hline
Purely Hadronic & $\frac{g_{\pi NN'}^2}{g_{\pi NN}^2}~<~10^{-2}$ & $\Gamma_{N'}~<~50~keV$ \\
                & $\frac{\sigma(pp\to nX^{++})}{\sigma(pn\to np)}~<~10^{-7}$\protect\cite{ram} & 
                  [$\frac{\Gamma_{N'}}{\Gamma_{\Delta}}~<~4\cdot 10^{-4}$] \\
                & $\frac{\sigma(pp\to\pi^+pX^0)}{\sigma(pp\to\pi^+pn)}~\sim 10^{-3}-10^{-4}$\protect\cite{Jiang} & \\
\hline
Hadronic and EM & $\frac{W(\pi^-p\to n'\gamma)}{W(\pi^-p\to n \gamma)}~<~8\cdot 10^{-5}~[\sim 10^{-5}]$ 
                & \\
                & $\Gamma_{N'\to N\gamma}~<~5~eV$\protect\cite{L'W} & $Br^2_{\gamma}~\Gamma_{p'}~<~10~eV$
                  \protect\cite{L'W} \\
                & $\frac{Y(ep\to e'\pi^+X^0)}{Y(ep\to e'\pi^+n)}~<~10^{-4}$\protect\cite{JLAB,MAMI} & 
                  $[\frac{Br_{\gamma}~\Gamma_{p'}}{Br_{\gamma}~\Gamma_{\Delta}}~<~2.8\cdot 10^{-3}$] \\
                & $\frac{Y(ed\to e'pX^0)}{Y(ed\to e'pn)}~<~10^{-4}$\protect\cite{MAMI} & \\
\hline
\end{tabular}
\end{center}
\end{table}
%%%%%%%%%%%%%%%%tbl.2
\begin{table}[h]
\caption{Possible unitary octet with $N'$ \label{tbl2}}
\begin{center}
\begin{tabular}{ccccc}
\hline
State     &  Mass        &   Width & Decay Modes       & Hadron \\
          &  (MeV)       &   (MeV) &                   & Production \\
          &              &         &                   & Cross Sections \\ 
\hline
N$'$      & $\sim$1100   & $<$0.05 & N$\gamma$ & $< 10^{-4}$ of ``normal" \\
$\Lambda$ & 1330         &         & $\Lambda\gamma$ & $\sim 10 \mu b$ \\  
$\Sigma$  & 1480         & 30-80   & $\Lambda\pi$, $\Sigma\pi$,N$\bar K$ & $\sim 1 \mu b$ \\
$\Xi$     & 1630         & 20-50   & $\Xi\pi$                            & $\sim  1 \mu b$ \\
\hline
\end{tabular}
\end{center} 
\end{table} 
%%%%%%%%%%%%%%%%%%%%%%%%%%%%%%%%%%%%%%%%%%%%%%%%%%%%%%%%%
%%%  Figures
%%%%%%%%%%%%%%%%%%%%%%%%%%%%%%%%%%%%%%%%%%%%%%%%%%%%%%%%%
\newpage
% === PSFIG 1 ======================================
% === PSFIG 2 ======================================
% === PSFIG 3 ======================================
% === PSFIG 4 ======================================
\newpage
% === PSFIG 5 ======================================
% === PSFIG 6 ======================================
% === PSFIG 7 ======================================
% === PSFIG 8 ======================================
% === PSFIG 9 ======================================
% === PSFIG 10 ======================================
% === PSFIG 11 ======================================
%================================================END

\end{document}